\documentclass{PoS}

\usepackage[utf8]{inputenc}
\usepackage[english]{babel}
\usepackage{amsmath}
\usepackage{amsfonts}
\usepackage{amssymb}
\usepackage{graphicx}
\usepackage{bm}
\usepackage{enumitem}
\usepackage{hyperref}
\usepackage{cite}

\newcommand{\be}{\begin{equation}}
\newcommand{\ee}{\end{equation}}
\newcommand{\bes}{\begin{equation*}}
\newcommand{\ees}{\end{equation*}}
\newcommand{\ba}{\begin{eqnarray}}
\newcommand{\ea}{\end{eqnarray}}

\newcommand{\bw}{\begin{widetext}}
\newcommand{\ew}{\end{widetext}}

\newcommand{\Imag}{\mathop{\mathrm{Im}}}

\title{Unitarized HEFT for strongly interacting longitudinal electroweak gauge bosons with resonances.}

\ShortTitle{Resonances in $W_LW_L$, $Z_LZ_L$ and $hh$ scattering}

\author{Rafael L. Delgado$^\dagger$, Antonio Dobado$^{*\&}$ and Felipe J. Llanes-Estrada$^*$\\
        $^\dagger$Physik-Department T30F, Technische Universit\"at M\"unchen, 85747 Garching, Germany\\
        $^*$Dept. F\'{\i}sica Te\'orica, Universidad Complutense, 28040 Madrid, Spain \\
$^\&$ Speaker and contact author}
\abstract{%
Higgs Effective Field Theory can be used to study vector-boson elastic scattering at the high energies relevant for the LHC. For most of the parameter space, the scattering is strongly interacting, with the minimal Standard Model being a remarkable exception. From its one-loop treatment complemented with dispersion relations and the Equivalence Theorem, we derive two  different unitarization methods which produce analytical amplitudes corresponding to different approximate solutions to the dispersion relations: the Inverse Amplitude method (IAM) and the N/D method.  
The partial waves obtained can show poles in the second Riemann sheet whose natural interpretation is that of dynamical resonances with masses and widths a function of the starting HEFT parameters. We compare the different unitarizations and we find that they are qualitatively, and in many cases quantitatively, very similar. However, for different reason it is more interesting to use one of the two methods depending on the particular channel for $WW$, $ZZ$, $WZ$, $Zh$, $Wh$ or $hh$ scattering. In this note we briefly describe the possible $I$ and $J$ channels for these reactions and give the unitarization method of choice in each case. The amplitudes obtained provide realistic resonant and nonresonant cross sections to be compared with and to be used for a proper interpretation of the LHC data.
}

\FullConference{The Sixth.  Annual Large Hadron Collider Physics conference LHCP 2018\\
		Sunday 3 -- Saturday 9, June 2018\\
		Bologna, Italy}

\begin{document}

The Higgs Effective Field Theory (HEFT) is an appropriate framework to describe a possible strongly interacting Symmetry Breaking Sector for the Standard Model (SM) as a non-linear effective field theory depending on several couplings which can be taken as free parameters. At the high LHC energies this strong dynamics could be probed through the longitudinal components of the $W$ and $Z$ bosons and the Higgs ($h$) interactions, possibly showing emergent dynamical resonances. With the Equivalence Theorem~\cite{Cornwall:1974km} (ET) one can obtain the high-energy amplitudes from those involving the would-be Goldstone bosons (WBGB) $\omega^a$ and the Higgs $h$, safe $O(M_W/\sqrt{ s})$ corrections. It is then consistent to neglect $M_W^2$, $M_Z^2$ and $M_h^2$, all around $(100\,{\rm GeV})^2$, against the $s$-scale $(1\,{\rm TeV})^2$. Then, the HEFT simplifies to the Next to Leading Order (NLO) effective Lagrangian~\cite{Buchalla:2015wfa,Kilian:2014zja,Delgado:2013loa,HEFT}:
\begin{eqnarray} \label{bosonLagrangian}
{\cal L} &=& \frac{1}{2}\left[1+2a\frac{h}{v}+b\left(\frac{h}{v}\right)^2\right]
\partial_\mu\omega^i\partial^\mu\omega^j\left(\delta_{ij}+\frac{\omega^i\omega^j}{v^2}\right)
+\frac{1}{2}\partial_\mu h\partial^\mu h %
\nonumber\\
 &+& \frac{4a_4}{v^4}\partial_\mu \omega^i\partial_\nu \omega^i\partial^\mu\omega^j\partial^\nu\omega^j
+\frac{4a_5}{v^4}\partial_\mu\omega^i\partial^\mu\omega^i\partial_\nu\omega^j\partial^\nu\omega^j
+\frac{g}{v^4}(\partial_\mu h\partial^\mu h)^2
\nonumber\\
 &+& \frac{2d}{v^4}\partial_\mu h\partial^\mu h\partial_\nu\omega^i\partial^\nu\omega^i
+\frac{2e}{v^4}\partial_\mu h\partial^\nu h\partial^\mu\omega^i\partial_\nu\omega^i
\ ,
\end{eqnarray}
which has seven parameters, $a$, $b$, $a_4$, $a_5$, $g$, $d$, and $e$. From this Lagrangian it is possible to compute for example the $\omega^a\omega^b \rightarrow \omega^c\omega^d$ NLO amplitudes (see the second entry in~\cite{Delgado:2013loa}) and the $IJ$ partial waves (with $I$ being the custodial isospin). Up to NLO, these partial waves take the general form
\begin{equation} \label{pertamplitude}
A_{IJ}^{(0)}(s) + A_{IJ}^{(1)}(s) = 
K s + \left(B(\mu)+D\log\frac{s}{\mu^2}+E\log\frac{-s}{\mu^2}\right)s^2.
\end{equation}
For physical $s$, perturbative unitarity relates $K$ and $E$ by $\Imag A^{(1)}_{IJ} = \lvert A^{(0)}_{IJ}\rvert^2$.  The $B(\mu)$ term contains the NLO low-energy constants that absorb the one-loop divergences and ensure order by order renormalizability. $K$ turns out to be  proportional to $(1-a^2)$. Similar expressions hold for the inelastic channel  $\omega\omega\to hh$ partial waves $M_J$ (and for the elastic $\omega  h\to\omega h, M'_J$), with $K$ proportional to $(a^2-b)$, and finally for the elastic channel $hh \to hh$ partial waves $T_J$ (in these, $K=0$). The complete formulae for the NLO $A_{IJ}$, $M_J$ and $T_J$ can be found in \cite{Delgado:PRD} (and in \cite{Axial} those for $M'_J$).

A nonvanishing Eq.~(\ref{pertamplitude}) signals a separation from the minimal SM. 
This happens as soon as $a=1=b^2$ is not satisfied, with tree-level amplitudes growing as $s$ indicating strong interactions (see the first reference in~\cite{Delgado:2013loa}). 
The NLO partial waves show the expected analytical structure on the $s$ complex plane (left and right cuts), but they are unitary only approximately and this for small energies. However, it is possible to complement the HEFT with dispersion relations~\cite{Truong:1988zp} so that the new amplitudes share the low-energy behavior of the HEFT, proper analytical structure and are also unitary. In addition, for some values of the couplings, they can show poles in the second Riemann sheet and close to the right cut that have the natural interpretation of dynamically generated resonances. This method is called \emph{Unitarized  Higgs Effective Field Theory} (UHEFT). However, dispersion relations are typically  integral equations of difficult solution. However, there are reliable algebraic approximation schemes. For example, the IAM solves very approximately the dispersion relation for the inverse of the partial waves. In the N/D, one starts from the perturbative result for $N$ and then solve a coupled system of integral equations by iteration. Usually,  both agree qualitatively, and in many cases also quantitatively, up to the first resonance. The differences come from the treatment of the amplitudes on the left-cut region. 

\begin{itemize} 
\item If there is only one channel (because $I\ne 0$ or $a^2=b$ decouple $\omega\omega$ and $hh$) one can split the NLO amplitudes $A^{(1)}$ into a left-cut and a right-cut carrying term and define a $g(s)$ function as:
\begin{equation}
\nonumber
 A_L(s) \equiv\left(\frac{B(\mu)}{D+E}+\log\frac{s}{\mu^2}\right) Ds^2\ ; 
\ \ \ 
A_R(s) \equiv\left(\frac{B(\mu)}{D+E}+\log\frac{-s}{\mu^2}\right) E s^2 \ ; 
\ \ \ 
g(s)\equiv\frac{1}{\pi}\left(\frac{B(\mu)}{D+E}+\log\frac{-s}{\mu^2}\right) .       
\end{equation}
In this left--right division the two pieces are, by construction, separately $\mu$ (renormalization scale)--independent. Then, as shown in~\cite{Delgado:PRD}, the elastic $\omega\omega$ partial waves, unitarized with the IAM and N/D method respectively are given by
\begin{eqnarray} \label{Atogether}
A^{IAM}(s) &=& \frac{\left[A^{(0)}(s)\right]^2}{A^{(0)}(s)-A^{(1)}(s)} 
 = \frac{A^{(0)}(s)+A_L(s)}{1-\frac{A_R(s)}{A^{(0)}(s)} -\left(\frac{A_L(s)}{A^{(0)}(s)}\right)^2 +g(s)A_L(s)};\nonumber\\
A^{N/D}(s) &=& \frac{A^{(0)}(s) + A_L(s)}{1-\frac{A_R(s)}{A^{(0)}(s)} +\frac{1}{2}g(s)A_L(-s)}. \nonumber
\end{eqnarray}
These two approximations to the partial waves have the proper analytical structure in the complex $s$ plane and are unitary. They are $\mu$--independent and inherit the low-energy behavior of the one-loop HEFT. In addition, for some regions of the coupling space they can show a pole in the second Riemann sheet at some point $s_0$. Such pole correspond to a dynamically generated resonance with mass $M$ and width $\Gamma$ given by $s_0=M^2-iM\Gamma$. In the cases where the right logarithm dominates over the left one, $A_R\gg A_L$, both unitarization methods yield the same resonances at approximately the same positions, since it is $A_L$ that causes a difference among them. In any case, for narrow resonances, $\gamma = \Gamma/M\ll 1$, 
the amplitude has a Breit-Wigner line shape with
\be
\nonumber
M^2= \frac{K}{B(M)}\ ;
\ \ \ \ 
\gamma= \frac{K^2}{B+D+E}. 
\ee
Reference~\cite{Espriu} illuminates the discussion by finding the regions of parameter space that support scalar, vector and isotensor resonances in the IAM method, and their corresponding masses and widths for $a^2= b$. 

\item
For two-channel amplitudes ($I = 0$ and $a^2 \ne b$), one has to consider the symmetric matrix
\be
F_{0J}(s)= %
\begin{pmatrix}
A_{0J}(s) & M_J(s) \\
M_J(s) & T_J(s) \\
\end{pmatrix}\ .
\ee
Now, unitarity requires that, on the right cut, $\Imag F(s) = F(s)F^\dagger(s)$, which is equivalent to
\be
 \nonumber  
 \Imag A_{0J} =  \lvert A_{0J}\rvert^2 + \lvert M_J\rvert^2 \ ;
 \ \ \ \ 
  \Imag M_J   =  A_{0J} M_{J}^*+  M_J T_J^* \ ;
  \ \ \ \ 
   \Imag T_J  =  \lvert M_J\rvert^2 + \lvert T_J\rvert^2\   .
\ee        
These relations do not hold exactly in NLO perturbation theory, since we only have
\be
\Imag F^{(1)}_{0J}(s) = F^{(0)}_{0J}(s)F^{(0)}_{0J}(s).  \nonumber
\ee
However, it is possible to solve this problem because both IAM and N/D methods can be extended to the coupled-channel case. The reason is that we are considering all particles as massless, so that 
the amplitudes share their analytical structure (cuts) avoiding the problems of overlapping cuts. Of course, as now we are dealing with matrices, attention must be paid to the order of the factors. For example, the IAM amplitudes now read
\be \label{IAMforF}
 F_{0J}^{\rm IAM}  =  F_{0J}^{(0)}(F_{0J}^{(0)}-F_{0J}^{(1)})^{-1}F_{0J}^{(0)},
\ee
where
\ba
F_{0J}(s) & = & F_{0J}^{(0)}(s)+F_{0J}^{(1)}(s)+\dots \nonumber \\     
F^{(0)}(s) & = &  K s\ ;
\ \ \ \ 
F^{(1)}(s)= \left(B(\mu) + D \log\frac{s}{\mu^2}+ E \log \frac{-s}{\mu^2} \right)s^2 \ ,
\label{pertmultichannel}
\ea 
and $K$, $B(\mu)$, $D$ and $E$ are $2\times 2$ matrices. Similarly, it is possible to generalize the N/D method to the $I=0$ ($a^2\ne b$) coupled channel case.  In Fig.~1 of~\cite{Delgado:PRD}, we show an example of how those methods can describe a isoscalar ($I=J=0$) resonance. It is important to realize that, in the coupled-channel IAM and the N/D method, the same poles appear in all the involved channels. It is also interesting to see that for some region of the couplings new genuine coupled channel resonances can appear which would not exist in absence of $\omega\omega$-$hh$ channel coupling (see the last reference in~\cite{Delgado:PRD}).
\end{itemize}

Now, since we have more than one unitarization method (the IAM, the N/D, and others not based in dispersion relations such as the improved K matrix), one may ask which is the most appropriate one if any. The answer depends on the considered channel. This is because of two reasons. First, the IAM methods cannot be applied if $K=0$. Second, the N/D cannot be applied in the case $D+E=0$. Thus, for the coupled isoscalar channels $I=J=0$, both methods can be applied giving similar results. (One could argue the superiority of the IAM, due to its simplicity, not requiring 
the $L$ and $R$ splitting of the NLO matrix element.) The elastic $\omega\omega$ and the inelastic $\omega h\to\omega h$ ($I=J=1$) cases (vector and axial channels) should be better unitarized using the elastic IAM method to avoid the instability in the region of the parameter space close to $D+E=0$ (as for example in a Higgsless QCD like theory). The inelastic $IJ=02$ has $K=0$ and then cannot be unitarized with the IAM method but can be unitarized with the coupled-channel N/D, which is stable in the limit $K=0$. Finally, for the elastic cases $I=2$ both for $J=0$ and $J=2$, also with $K=0$, one can use the simpler elastic $N/D$ method too. In this way, it is possible to unitarize all non vanishing channels appearing in the HEFT up to the one-loop level.

In conclusion, the UHEFT, constructed from the NLO HEFT results and  dispersion relations solved (approximately) by the IAM and the N/D methods, can provide a down-top description of a general strongly interacting symmetry breaking sector of the SM (under intense discussion in the context of composite scenarios). This UHEFT can also describe dynamically generated resonances for the different channels with properties depending on the effective Lagrangian couplings. The corresponding amplitudes can be used for a phenomenological description of the resonant (or not) LHC data. See, for instance, Ref.~\cite{Vector} for the case of vector resonances. The UHEFT can also be extended to encompass $\gamma\gamma$ initial or final states~\cite{Delgado:2016rtd} and to final $t\bar t$ states~\cite{Castillo:2016erh}.

\begin{figure}
\includegraphics[width=0.48\textwidth]{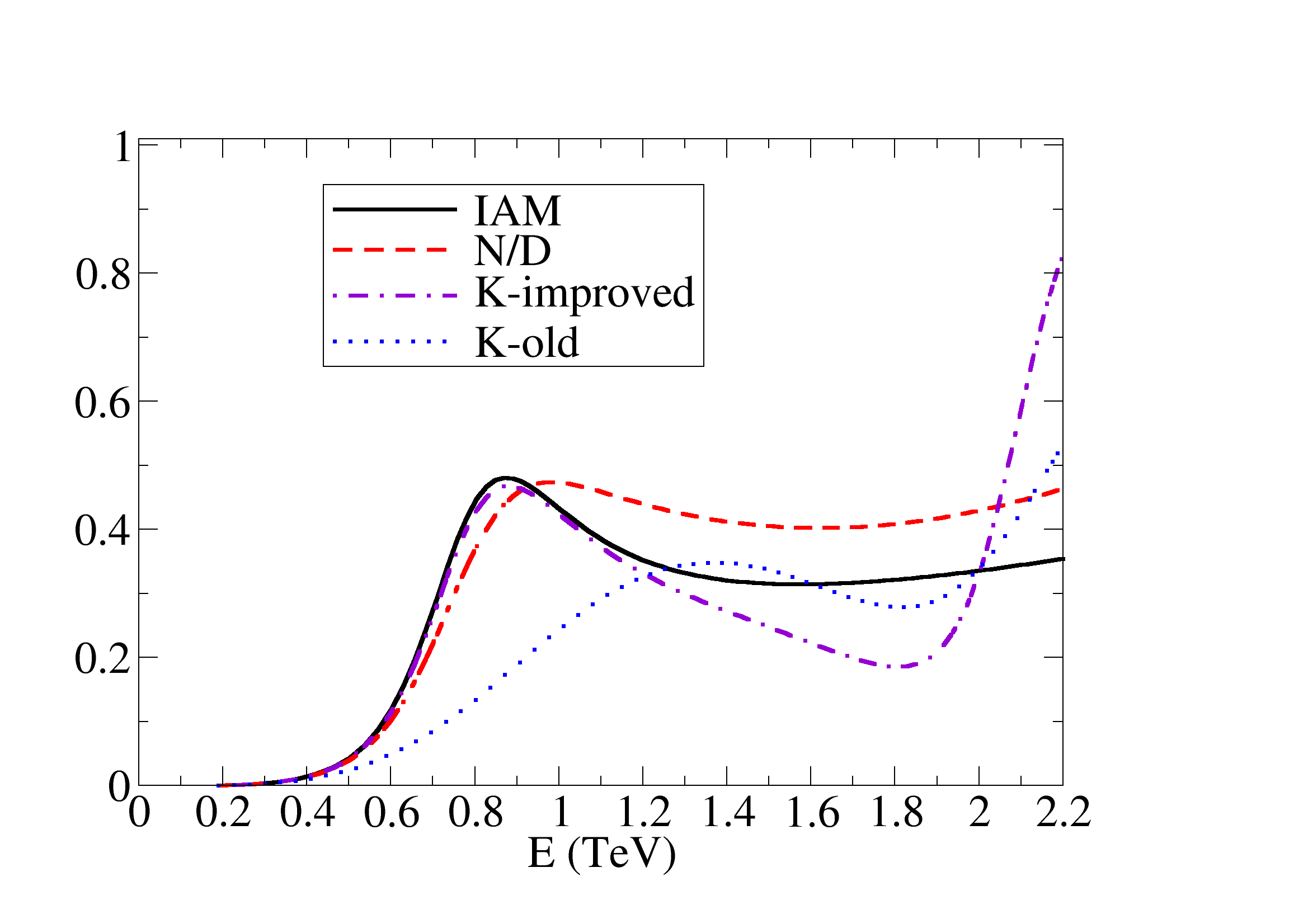}
\includegraphics[width=0.48\textwidth]{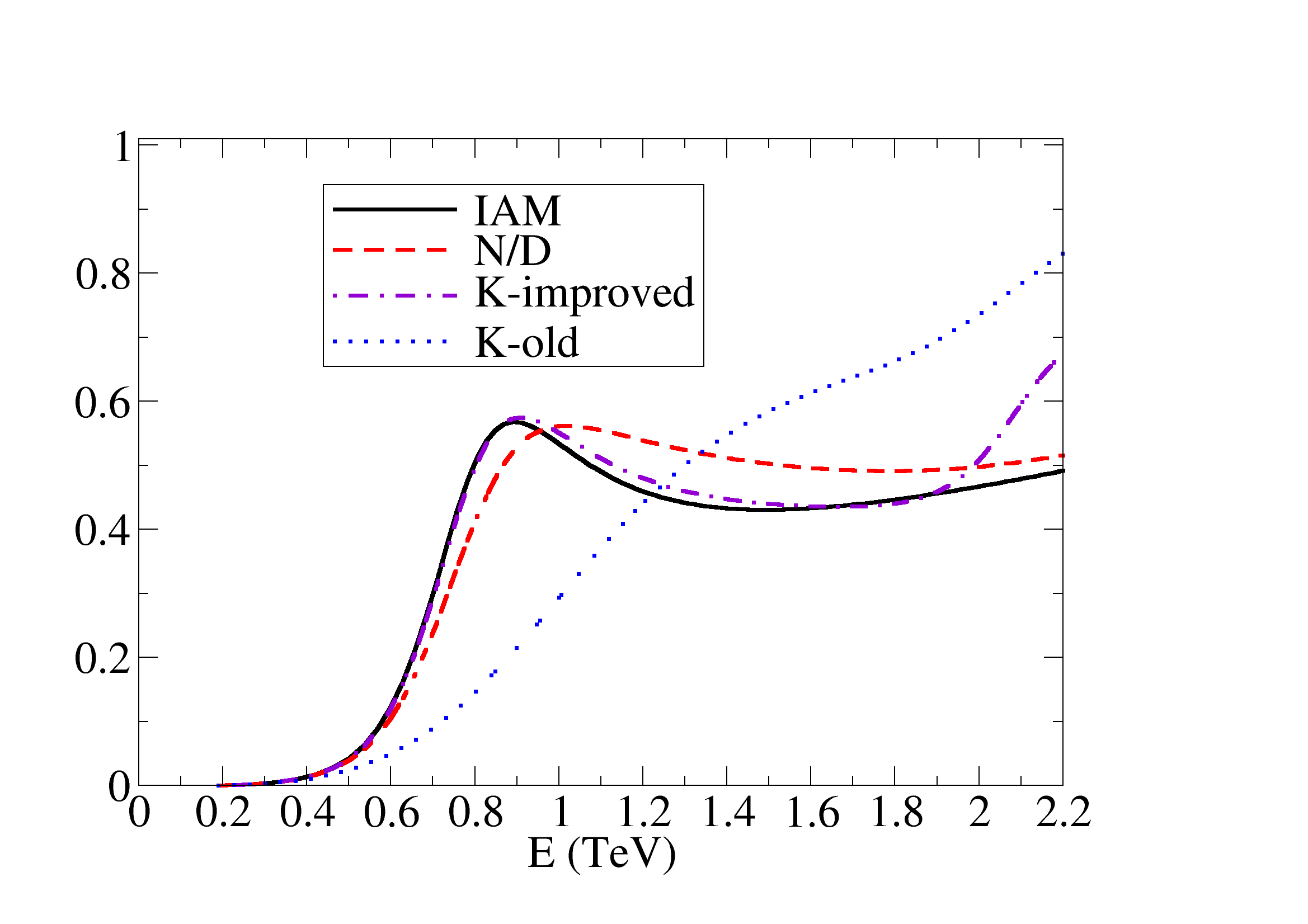} \\
\begin{minipage}{0.48\textwidth}
\includegraphics[width=0.95\textwidth]{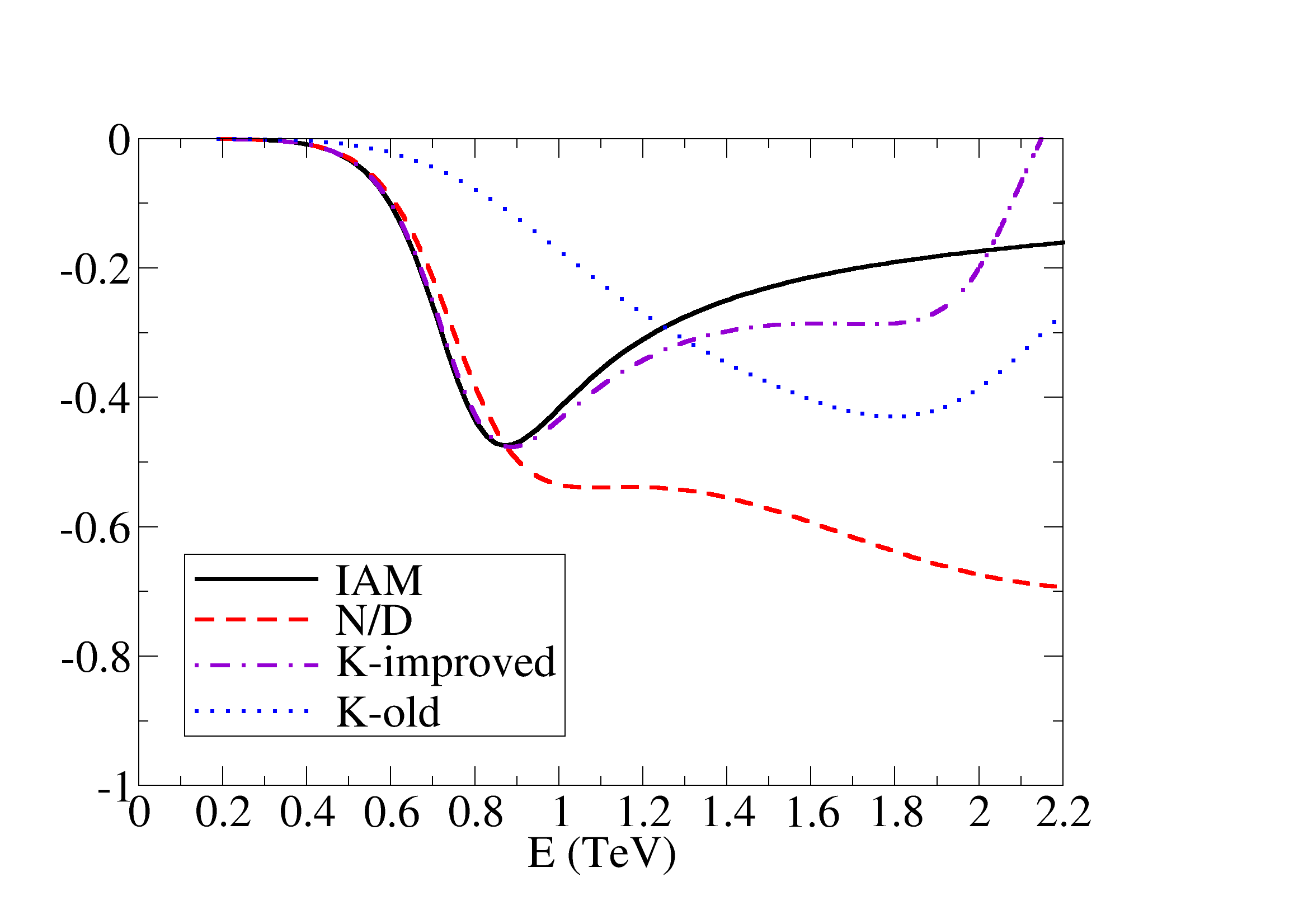}
\end{minipage}
\begin{minipage}{0.48\textwidth}
\caption{\label{fig:IAM} We compare different unitarization methods
for the imaginary parts of the $IJ=00$ amplitudes. Clockwise from top left, $\omega\omega$, $hh$ and cross-channel $\omega\omega\to hh$ (with LO parameters $a=0.88$ and $b=3$, $\mu=3\,{\rm TeV}$ and the NLO ones set to zero). A scalar resonance is visible in all, and the IAM and N/D closely agree. We consider also the (old) K matrix method, and the so called improved K matrix method (which is analytical and unitary as the IAM and the N/D) (see~\cite{Delgado:PRD}).}
\end{minipage}
\end{figure}

We thank J. J. Sanz Cillero and  D. Espriu for collaboration and useful discussions. Work supported by the Spanish grant FPA2016-75654-C2-1-P (A. D. and F. J. L.-E.) and the "Ram\'on Areces" Foundation (R. D.).

\newpage

\end{document}